\documentclass[11pt]{article}

\usepackage{graphicx}

\usepackage{amsfonts}
\usepackage{amsmath}

\usepackage{framed}

\usepackage{amsthm}

\newlength{\defbaselineskip}
\begin{document}

\title{Neuron as a reward-modulated combinatorial switch and a model of learning behavior}
\author{Marat M. Rvachev\footnote{Email: rvachev@alum.mit.edu}}

\date{April 27, 2013}

\maketitle

\begin{abstract}
This paper proposes a neuronal circuitry layout and synaptic plasticity principles that allow the (pyramidal) neuron to act as a "combinatorial switch". Namely, the neuron learns to be more prone to generate spikes given those combinations of firing input neurons for which a previous spiking of the neuron had been followed by a positive global reward signal. The reward signal may be mediated by certain modulatory hormones or neurotransmitters, e.g., the dopamine. More generally, a trial-and-error learning paradigm is suggested in which a global reward signal triggers long-term enhancement or weakening of a neuron's spiking response to the preceding neuronal input firing pattern. Thus, rewards provide a feedback pathway that informs neurons whether their spiking was beneficial or detrimental for a particular input combination. The neuron's ability to discern specific combinations of firing input neurons is achieved through a random or predetermined spatial distribution of input synapses on dendrites that creates synaptic clusters that represent various permutations of input neurons. The corresponding dendritic segments, or the enclosed individual spines, are capable of being particularly excited, due to local sigmoidal thresholding involving voltage-gated channel conductances, if the segment's excitatory and absence of inhibitory inputs are temporally coincident. Such nonlinear excitation corresponds to a particular firing combination of input neurons, and it is posited that the excitation strength encodes the combinatorial memory and is regulated by long-term plasticity mechanisms. It is also suggested that the spine calcium influx that may result from the spatiotemporal synaptic input coincidence may cause the spine head actin filaments to undergo mechanical (muscle-like) contraction, with the ensuing cytoskeletal deformation transmitted to the axon initial segment where it may modulate the global neuron firing threshold. The tasks of pattern classification and generalization are discussed within the presented framework.
\end{abstract}

\section{Introduction}
\label{sec1}
The field of reinforcement learning (RL) solves the problem of sequential decision making by an agent receiving delayed numerical rewards \cite{sutt98}. The field can be viewed as originating from two major threads: the idea of learning by trial and error that started in the psychology of animal learning (e.g., \cite{thor11}), and the problem of optimal control and its solution using value functions and dynamic programming \cite{bell57}. An important branch of the RL theory is the temporal difference (TD) class models for the phasic activity of midbrain dopamine neurons \cite{mont95,mont96,schu97a}. The dopamine activity is believed to encode a reward prediction error (RPE) signal that guides learning in the frontal cortex and the basal ganglia \cite{bush51a,bush51b,schu97,schu06}. Most scholars active in dopamine studies believe that the dopamine signal adjusts synaptic strengths in a quantitative manner until the subject's estimate of the value of current and future events is accurately encoded in the frontal cortex and basal ganglia \cite{glim11}. 

This paper considers the problem of instantaneous decision making by an agent receiving immediate rewards within an RL-type framework. A trial-and-error learning paradigm is suggested in which the reward signal modulates memory in (cortical) neurons that act as combinatorial switches. The reward signal may come from an "elementary" reward generator such as that reflecting pain or satisfaction of hunger; it may also involve an RPE-type or "critic"-type~\cite{sutt98} signal mediated by dopamine and/or other agents that could convey positive as well as negative reward components as was first suggested in \cite{daw02}. 

The first contributing thread to the presented model, as in the classical RL theory, is the idea of learning by trial and error and reinforcement of favorable outcomes. The idea, as expressed in Edward Thorndike's "Law of Effect" \cite{thor11}, is: "Of several responses made to the same situation those which are accompanied or closely followed by satisfaction to the animal will, other things being equal, be more firmly connected with the situation, so that, when it recurs, they will be more likely to recur; those which are accompanied or closely followed by discomfort to the animal will, other things being equal, have their connections to the situation weakened, so that, when it recurs, they will be less likely to occur. The greater the satisfaction or discomfort, the greater the strengthening or weakening of the bond." This idea is widely regarded as a basic principle underlying much behavior \cite{hilg75,denn81,camp60,czik95}. 

The second contributing thread is a novel idea that, given proper neuronal circuitry layout, pyramidal neurons can process information by switching the neuron output based on active input neuron combinations. This  idea builds on the Two-Layer Neural Network (TLNN) model for the pyramidal neuron~\cite{p03}. Additional computational advantages that could make the idea possible may be provided by mechanical force generated at the dendritic spines and stretch-activation of Na$^+$ channels at the axon initial segment. An interesting feature of the presented framework is its ability to distil reusable abstract concepts about the environment, making learning with the low-dimensional feedback signal, the reward, efficient. 

\subsection{Problem formulation}
\label{sec1.1}
\begin{figure}[t]
\begin{center}
\includegraphics[width=\textwidth]{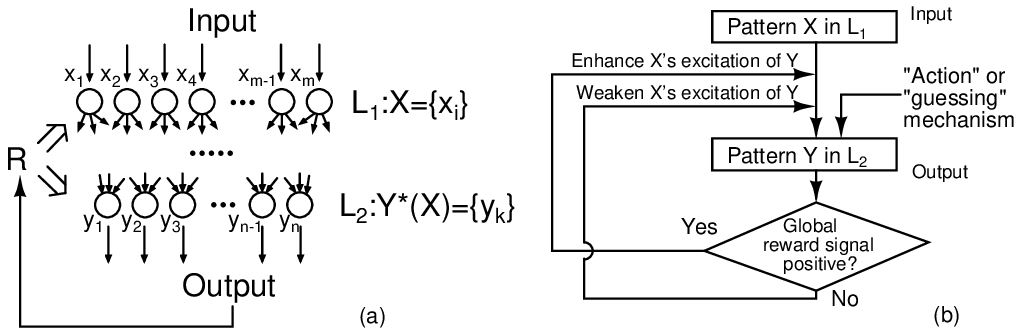}
\caption{\label{fig1} The organism-level learning problem and an outline of the suggested solution. (a) Formulation of the problem. Neurons $x_{i}$, $i=1,\ldots,m$ in layer $L_1$ connect to neurons $y_k$, $k=1,\ldots,n$ in layer $L_2$. A pattern of excitations $X=\{x_{i}\}$, if responded to by a pattern of excitations $Y=\{y_k\}$, elicits a positive or negative reward $R$ resulting from the interaction of the generated motor behavior with the environment. The problem is: given an arbitrary $X$, excite $Y^*(X)$ that would lead to positive $R$. (b) Outline of the suggested solution. Learning proceeds by trial and error. Excitation of pattern $X$ excites a pattern $Y(X)$, possibly with the help of an "action" mechanism (e.g., depolarization to all $L_2$ neurons until a certain level of the aggregate $L_2$ output activity is achieved, as discussed in Sec.~\ref{sec2.3}). A "guessing" mechanism introduces variations in the excited patterns $Y$. $X$'s excitation of those $Y$ that lead to positive (negative) $R$ is enhanced (weakened).}
\end{center}
\end{figure}

The following organism-level learning problem is posed. For simplicity, the neuronal activity states are considered to be binary: "firing" or "not firing". Given an arbitrary combination $X$ of firing neurons in a (perhaps sensory) input layer $L_1$, activate a corresponding "optimal" combination $Y^*(X)$ of firing neurons in a (perhaps motor) output layer $L_2$ (Fig.~\ref{fig1}(a)). The optimal combination $Y^*(X)$ is defined as one that produces the motor behavior that results in a positive global reward signal $R$ in the organism. As such, $Y^*(X)$ can be an arbitrary combination of $L_2$ neurons from a combinatorics perspective. The reward signal $R$, in biological terms, may be mediated by certain modulatory neurotransmitters or hormones that are diffusely delivered to generally trainable neurons. It is assumed that in biological systems $R$ can be activated by evolutionarily hardwired circuits, such as when hunger is satisfied, as well as by higher mental processes, e.g., due to the organisms' subjective evaluation of the motor behavior as being satisfactory given the sensory inputs.

\begin{figure}[]
\begin{center}
\includegraphics[width=0.45\columnwidth]{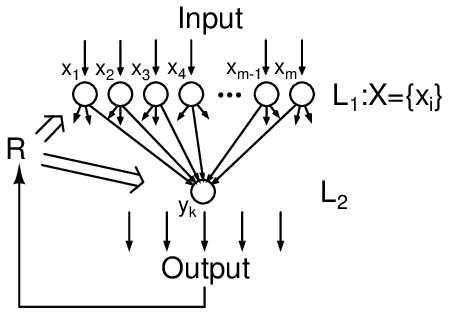}
\end{center}
\caption{\label{fig2} The single-neuron learning problem. $L_1$ neurons $x_i$ connect to an $L_2$ neuron $y_k$. Long-term enhance (weaken) $y_k$ excitation by those combinations $X$ for which the following $y_k$ excitation resulted in a positive (negative) $R$. The enhancement and weakening of excitation may involve long-term potentiation (LTP) and long-term depression (LTD) processes that are influenced by both the combinatorics of the problem and the reward $R$, as suggested in Sec.~\ref{sec2.2.1}.}
\end{figure}

It is suggested that the learning process proceeds in a trial-and-error fashion. Given a firing combination $X$ variations are introduced in the firing combination $Y$ with the $X$'s excitation of those $Y$ that lead to positive $R$ being enhanced while $X$'s excitation of those $Y$ that lead to negative $R$ being weakened (Fig.~\ref{fig1}(b)). Details of this suggested process are discussed in more detail in Sec.~\ref{sec5}. First, a more elementary learning task is considered: given an arbitrary firing combination $X$ long-term strengthen excitation of an $L_2$ neuron $y_k$, specifically by $X$, if the subsequent reward $R$ is positive. Conversely, long-term weaken excitation of $y_k$, specifically by $X$, if $R$ is negative (Fig.~\ref{fig2}).

\section{Solution to the single-neuron combinatorial switching problem}
\label{sec2}
\subsection{Local dendritic integration as the basis for combinatorial memory}
\label{sec2.1}
The following mechanism is posited as the solution and is illustrated in Figs. 3 and 4. 
\begin{figure}[t]
\begin{center}
\includegraphics[width=0.83\columnwidth]{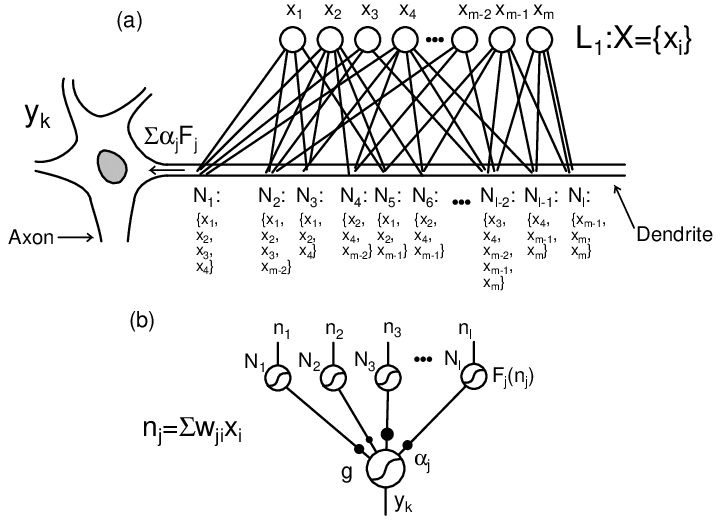}
\end{center}
\caption{\label{fig3} Solution to the single-neuron learning problem shown in Fig.~\ref{fig2}. (a) Inputs from $x_i$ form spatially localized clusters, or "neighborhoods", $N_j$, $j=1,\ldots,l$, on the $y_k$ dendrites. In the figure, a set of $x_i$ below an $N_j$ denotes neurons projecting into the cluster. Activation of excitatory synapses simultaneous with a lack of activation of inhibitory synapses in cluster $N_j$ produces output $F_j$ that has a "combinatorial memory" component $C_j$. $F_j$ generated at all $y_k$ neighborhoods superimpose.  $C_j$ expression is regulated as suggested in Fig.~\ref{fig4}. (b) An equivalent neural network diagram for (a). The synaptic weight from $x_i$ to $N_j$ is $w_{ji}$. Neighborhood weights $\alpha_j$ are shown in filled circles. Both figures (a) and (b) are essentially a reinterpretation of Fig.~1 in~\cite{p03}.}
\end{figure}
\begin{figure}[t]
\begin{center}
\includegraphics[width=0.55\columnwidth]{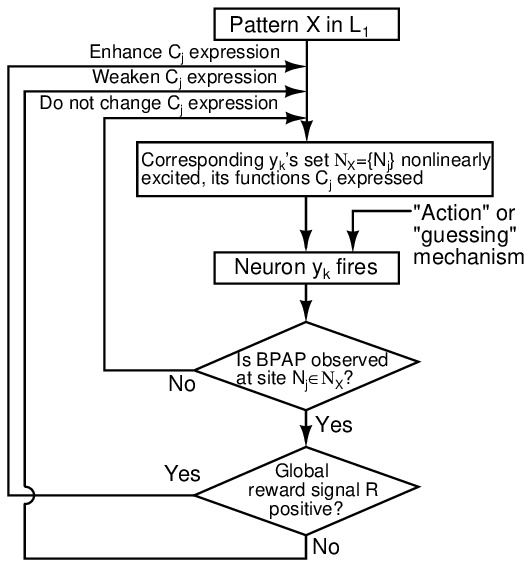}
\end{center}
\caption{\label{fig4} Suggested learning rules for the combinatorial memory $C_j$. The excitation of a pattern $X$ in $L_1$ leads to the excitation of the corresponding set of neighborhoods $\mathbb{N}_X=\{N_j\}$ in $y_k$. The input-output function $C_j$ is long-term enhanced (weakened) if the neighborhood $N_j$ is excited, i.e., $N_j\in \mathbb{N}_X$, this is closely followed by a BPAP at $N_j$ and the following $R$ is positive (negative).}
\end{figure}
$L_1$ neurons connect to the $y_k$ dendrites at random or predetermined locations, forming spatially localized (and possibly overlapping) "synapse neighborhoods" $N_j$ that contain various permutations of input neurons. Sufficient depolarization of the dendritic and/or spine interior within the $j$th neighborhood, caused by the temporal coincidence of the neighborhood's excitatory and absence of inhibitory inputs, causes $N_j$ excitation. The $N_j$ excitation drives local input-output function $F_j$ that has a "combinatorial memory" input-output component $C_j$ that possesses the following properties: 1) $C_j$ expression is long-term enhanced (weakened) if the neighborhood $N_j$ is excited, this is closely followed by a back-propagating action potential (BPAP) at $N_j$, and the immediately following $R$ is positive (negative), and 2) compared to other drivers of neuron stimulation, $C_j$ can substantially contribute to the $y_k$ excitation. Note that the input-output function $C_j$ is driven by $N_j$ excitation that itself is caused by the spatiotemporal coincidence of inputs. This confers $C_j$ combinatorial specificity.

As an example, assuming, as we do throughout this paper, that all inputs $\{x_i\}$ are 1 or 0, i.e., active or inactive, and also that all synaptic weights $w_{ji}$ from $x_i$ to $N_j$ are +1, -1 or 0 (corresponding to excitatory, inhibitory synapses and the absence of synaptic contact, respectively), a simple $C_j$ can be written as
\begin{equation}
\label{eq1}
C_j(X) = \gamma_j H(n_j-n_j^*-\overline{n}_j),
\end{equation}
where $\gamma_j$ is the weight on $N_j$,  $n_j^*=\sum_{w_{ji}>0}w_{ji}>0$ is the number of $N_j$ excitatory synapses,  $n_j=\sum_{w_{ji}>0}w_{ji}x_i$ is the number of active $N_j$ excitatory synapses ($n_j^* \ge n_j \ge 0$), $\overline{n}_j=-\sum_{w_{ji}<0}w_{ji}x_i \ge 0$ is the number of active $N_j$ inhibitory synapses and $H(n)$ is the step function:
\begin{equation}
\label{eq2}
H(n)= \left\{
\begin{array}{ll}
      1 & \quad  \textrm{if $n\geq 0$},\\
      0 & \quad \textrm{if $n < 0$}. \\
\end{array} \right.
\end{equation}
In Eq.~\ref{eq1} the argument to $H()$ is less than 0 unless 1) $n_j^*  = n_j$, i.e., all excitatory synapses in $N_j$ are active and 2) $\overline{n}_j = 0$, i.e., all inhibitory synapses in $N_j$ are inactive (recall that $n_j^* > 0$, $n_j^* \ge n_j \ge 0$ and  $\overline{n}_j \ge 0$). Weights $\gamma_j$ are increased (decreased) if $N_j$ is excited, this is closely followed by a BPAP at $N_j$ and the immediately following $R$ is positive (negative). Note that in this formulation  weights $\gamma_j$ are independent from $w_{ji}$ and learning may proceed with changing $\gamma_j$ and unchanged $w_{ji}$.

It is easy to see that the existence of the input-output functions $C_j$ can in principle solve the single-neuron learning problem posed in Fig.~\ref{fig2}, assuming that $y_k$ firings cause BPAPs that propagate to all $N_j$ (we do assume this here and below). Indeed, insofar as the features of $X$ are represented in the corresponding set of excited neighborhoods $\mathbb{N}_X=\{N_j\}$, the corresponding  set $\mathbb{C}_X=\{C_j\}$ will be strengthened and will enhance $y_k$ excitation when $X$ is presented, if an earlier $X$ presentation was followed by $y_k$ firing and the subsequent $R$ was positive. This immediately follows from the $C_j$ training rules (Fig.~\ref{fig4}). The combinatorial specificity of $C_j$ should ensure that the enhanced $y_k$ excitation will be specific to the pattern $X$ and similar patterns (see Sections \ref{sec3} and \ref{sec6} for numerical simulations). Conversely, $C_j$ training rules will cause a weakened $y_k$ excitation by those $X$ for which a previous $y_k$ excitation was followed by negative $R$. In Sec.~\ref{sec2.2} we digress into looking for a plausible physical (electrical or mechanoelectrical) $C_j$ realization in biological neurons, and starting in Sec.~\ref{sec2.3} we model networks of neurons possessing $C_j$.

\subsection{Possible physical realizations of the combinatorial memory}
\label{sec2.2}
\subsubsection{Electrical mechanism}
\label{sec2.2.1}
\begin{figure}[t]
\begin{center}
\begin{tabular}{cc}
\includegraphics[width=.5\textwidth]{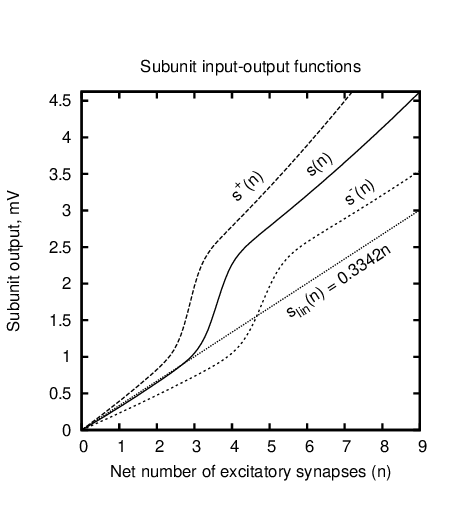} &
\includegraphics[width=.5\textwidth]{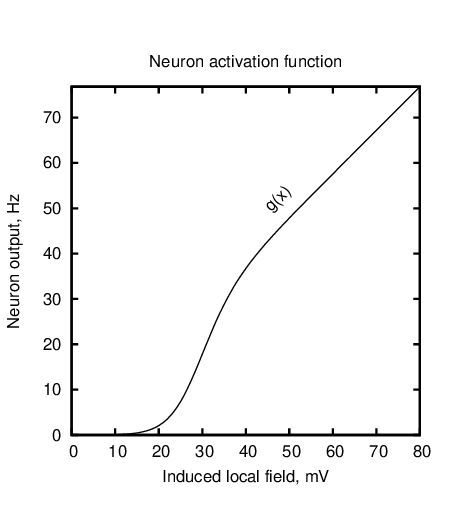} \\
  \end{tabular}
\end{center}
\caption{\label{figureSnScaled} The subunit input-output function from \cite{p03}: $s(n)=1/(1+exp((3.6-n)/0.2) ) +0.3 n + 0.0114 n^2$. Also shown are estimated subunit input-output functions after a uniform 25\% increase and decrease in efficacy of all subunit synapses: $s^+(n) = s(1.25n)$ and $s^-(n) = s(0.75n)$. The function $s_{lin}(n)=0.3342n$ is the linear fit to $s(n)$ below $n=3$.  (b) The neuron global activation function from \cite{p03}: $g(x) = 0.96 x/(1+1509 exp(-0.26x))$.}
\end{figure}

The local input coincidence detection that confers $C_j$ the combinatorial nature can be related to the local sigmoidal thresholding of postsynaptic potentials (PSPs) that results from nonlinear activation of voltage-dependent NMDA, Ca$^{2+}$ and Na$^{2+}$ currents \cite{pols04,silv10,haus03,p03}. Using a detailed compartmental model of a hippocampal CA1 pyramidal neuron, the neuron input-output function was shown to behave as a two-layer "neural network" with the output given by \cite{p03}
\begin{equation}
\label{eq3}
y= g\left[\sum_{j=1}^l\alpha_j s(n_j)\right],
\end{equation}
where $n_j$ is the net number of excitatory synapses driving the $j$th dendritic "subunit", $s(n)$ is the subunit input-output function, $\alpha_j$ is the weight on the $j$th subunit, $l$ is the number of subunits in the cell, and $g(x)$ is the global output nonlinearity (Fig.~\ref{figureSnScaled}). 
In the study the subunits were assumed to correspond physically to long, thin unbranched terminal dendrites of the apical and basal tree. The strength of each synapse was scaled to yield an equal (5 mV) peak EPSP locally at each synapse for the input intensities simulated. We estimate the effect of a uniform 25\% increase and decrease in efficacy of all subunit synapses on the subunit input-output function in Fig.~\ref{figureSnScaled}(a), assuming that the scaling in synaptic strength is equivalent to the scaling in the number of active synapses. 

\begin{figure}[t]
\begin{center}
 \includegraphics[width=.5\textwidth]{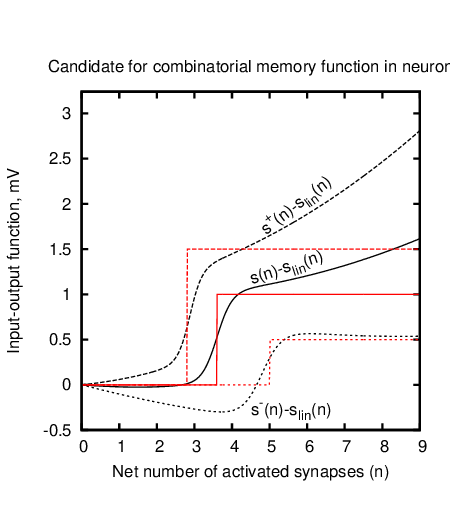} 
\end{center}
\caption{\label{figSnScaledDiff} Estimated effect of a uniform 25\% increase and decrease in efficacy of all subunit synapses on the function $C(n) = s(n) - s_{lin}(n)$ (black lines). Plotted in red is a similar hypothetical combinatorial memory function $C^*(n) = \gamma H(n-n_{tr})$ with a trainable weight $\gamma$ and activation threshold $n_{tr}$.}
\end{figure}

Lets suppose the subunit's "combinatorial memory" input-output component is given by $C(n) = s(n) - s_{lin}(n)$, where $s_{lin}(n) = 0.3342n$ is the linear fit to $s(n)$ below the threshold (Fig.~\ref{figureSnScaled}(a)). Indeed, as shown in Fig.~\ref{figSnScaledDiff}, 
$C(n)$ is increased (decreased) and its "activation threshold" is lowered (raised) following a uniform increase (decrease) in the synapse efficacy, making $C(n)$ a good candidate for a trainable detector of local input coincidences if the linear component $ s_{lin}(n)$ can be subtracted from $s(n)$ in the signal analysis. In fact, it will be shown in Sec.~\ref{sec3.2} that the lowered $C(n)$ activation threshold when the synaptic efficacy is increased is not likely to improve the neuron performance in the task of generalization. However, in order for $C(n)$ to act as the combinatorial memory defined in Sec.~\ref{sec2.1}, the individual synapse plasticity has to be influenced by both the combinatorics of the problem and the reward $R$. For example, $C(n)$ will conform to the combinatorial memory plasticity rule if the synaptic plasticity behaves as: LTP (LTD) is induced in a synapse if the synapse and its subunit are excited, this is immediately followed by a BPAP at the subunit and the immediately following $R$ is positive (negative). 

\subsubsection{Mechanical (muscle-like) mechanism}
\label{sec2.2.2}
\begin{figure}[t]
\begin{center}
\includegraphics[width=0.8\columnwidth]{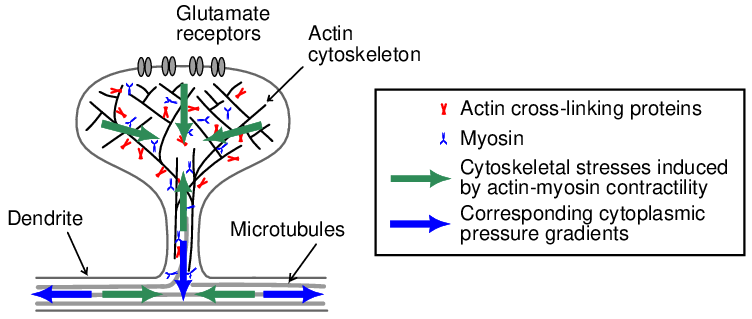}
\end{center}
\caption{\label{figSpine} The influx of Ca$^{2+}$ into a spine could elicit (muscle-like) contraction in the spine actin cytoskeleton. Induced cytoskeletal stresses and cytoplasmic pressure gradients are shown as green and blue arrows, respectively. These mechanical forces could propagate along the dendrite, as shown. At the axon initial segment, the ensuing stresses could modulate the global neuron firing threshold.}
\end{figure}

An interesting $C_j$ realization is possible if the free calcium entering a spine during a spatiotemporal synaptic input coincidence event elicits spine actin filament contraction, e.g., through calcium-activated actin interaction with myosin or another actin-binding protein. The ensuing cytoskeletal and cytoplasmic stresses, the magnitudes of which could encode the combinatorial memory, could be transmitted along the dendritic shaft to the $y_k$'s axon initial segment (Fig.~\ref{figSpine}).
At the initial segment these stresses, superimposed with those generated at other dendritic sites, could regulate the global $y_k$ excitation threshold via stretch-modulating Na$^+$ voltage-gated ion channels (Nav)~\cite{rva03,rvac10}. The use of the mechanical force would provide the second dimension to the neuron's computational machinery, disentangling the spatiotemporal coincidence detection mechanism, which would be electrical and based on local nonlinear voltage summation, from the $C_j$ readout mechanism, which would be mechanical. The spine head volume and the associated quantity of actin filaments would then reflect the $C_j$ magnitude, rather similarly to how the muscle cell volume and strength reflect the memory of previous exercise.

\begin{figure}[t]
\begin{center}
\includegraphics[width=.5\textwidth]{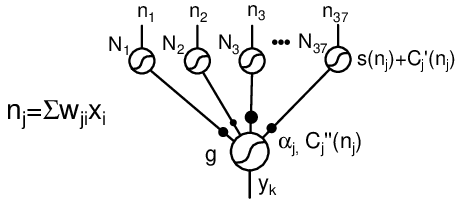} 
\end{center}
\caption{\label{figMechSimLayout} Equivalent neural network for simulation of a pyramidal neuron with hypothetical mechanical memory. Compared to the TLNN model~\cite{p03} expressed by Eq.~\ref{eq3}, each subunit has an additional "local" mechanical memory output $C'_j$ as well as it contributes $C''_j$ to the global nonlinearity input.}
\end{figure}

The advantage of having additional mechanical memory can be seen from the following simulation. Using simulation results from \cite{p03}, we modeled a pyramidal neuron with 37 dendrites (the "subunits", or "branches") connecting to the apical trunk (Fig.~\ref{figMechSimLayout}). 
As in \cite{p03}, the neuron inputs were excitatory only. The dendrite input-output functions $s(n)=1/(1+exp((3.6-n)/0.2) ) +0.3 n + 0.0114 n^2$ and the global output nonlinearity $g(x) = 0.96 x/(1+1509 exp(-0.26x))$ that in \cite{p03} were combined using Eq.~\ref{eq3}, were modified with two versions of "mechanical memory", $C'_j(n_j) = \gamma_j' H(n_j-4)$ and $C_j''(n_j) = \gamma_j'' H(n_j-4)$, as
\begin{equation}
\label{eq4}
S_j(n_j) = s(n_j)+ C_j'(n_j), \quad j=1, \ldots, 37,
\end{equation}
\begin{equation}
\label{eq5}
y= g\left[ x_0 +  \sum_{j=1}^{37}\left[ \alpha_j S_j(n_j)+C''_j(n_j) \right]   \right],
\end{equation}
where $n_j$ is the net number of active inputs in subunit $j$ and $\alpha_j$, $\gamma_j'$, $\gamma_j''$ and $x_0$ are constants. The function $C'_j(n_j)$ represented an added output invoked in the subunit $j$ by a local mechanical memory mechanism (e.g., the local actin cytoskeleton contractility stretch-modulating gating of the membrane ion channels). The function $C''_j(n_j)$ represented modulation of the $y_k$ firing threshold by a global mechanical memory mechanism. Both mechanisms were activated when $n_j$ was at least four. The task posed for the neuron was to detect coincident activation of at least 4 inputs on any of the branches, signaling it by firing with the frequency at least 40 Hz. The coincidence threshold was set at 4 because the electrical subunit functions $s(n)$ display a sharp increase at $n=4$ (Fig.~\ref{figureSnScaled}(a)) which should help the neuron detect such patterns. For the purely electrical neuron ($\gamma'_j=\gamma''_j=0$) the global output $g(x)$ threshold was allowed to vary via the parameter $x_0$ to improve the detection performance.

The total of 7000 input patterns were generated, 3500 with four active synapses on a branch (the coincidence pattern), and 3500 with at most three active synapses on a branch (the no-coincidence pattern). To sample a wide range of input intensities, patterns were selected as follows. First, the total number of active synapses $N_e$ was chosen at 30, 35, 40, 45, 50, 55 or 60. Five no-coincidence patterns were generated by  randomly distributing $N_e$ active inputs on the 37 branches, restricting the number of active inputs to at most three per branch. One coincidence pattern was generated for each $c$ from 1 to 5, by assigning four active inputs to $c$ random branches; the remaining ($N_e-4c$) active inputs were randomly distributed to the remaining ($37-c$) branches, restricting the number of active inputs per branch to at most three. The above procedure for generating 10 patterns was repeated 100 times for each $N_e$, yielding 7000 patterns overall (7 x 10 x 100).

\begin{figure}[t]
\begin{center}
\begin{tabular}{cc}
    \includegraphics[width=.5\textwidth]{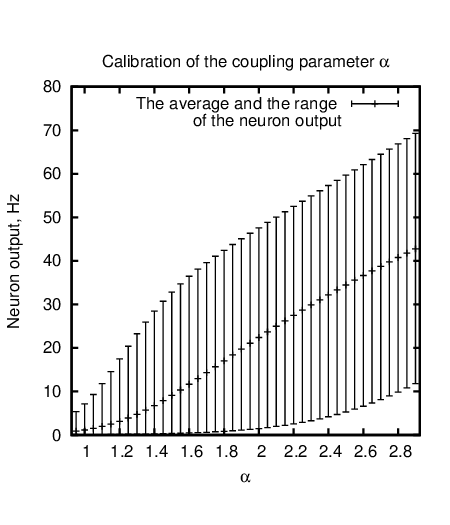} &
   \includegraphics[width=.5\textwidth]{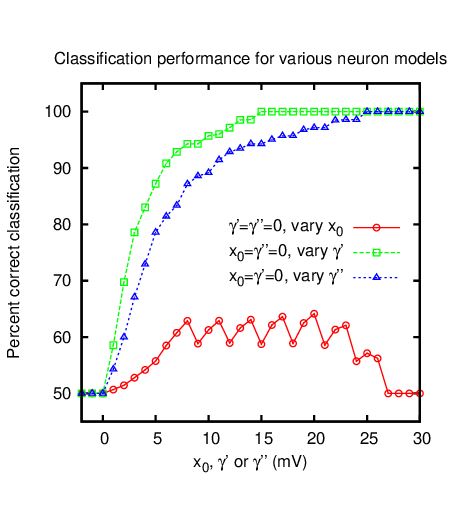}   \\
\end{tabular}
\end{center}
\caption{\label{mechSim} (a) The average and the range of output for the electrical neuron ($\gamma'_j=\gamma''_j=0$) for the 7000 input patterns, as a  function of $\alpha=\alpha_i$. The error bars show the minimum and maximum neuron output over all the input patterns. (b) Classification performance for the 7000 patterns for the three neuron models: purely electrical (red circles), with local mechanical memory (green squares) and with global mechanical memory (blue triangles). In all cases $\alpha=\alpha_i = 1.7$. The "mechanical memory" weights for the 37 subunits were assumed equal: $\gamma'=\gamma'_j$, $\gamma''=\gamma''_j$, $j=1,\ldots , 37$. No parameter fine tuning was required to obtain the results shown other than setting $\alpha$ to 1.7. For $\alpha$ equal to 1 (2.5) the best classification performance was 63\%, 100\%, 100\% (64\%, 86\%, 86\%) for the electrical and the two mechanical neurons, respectively.}
\end{figure}

To simplify, all $\alpha_i$ were assumed to be equal and were scaled in a pilot run with $\gamma'_j = \gamma''_j = 0$ so that a reasonable sub-40-Hz output was observed for the 7000 patterns, giving $\alpha=\alpha_i=1.7$ (Fig.~\ref{mechSim}(a)). 
Note that in the more realistic neuron model \cite{p03}, the ratio of the average couplings  $\alpha_i$ for the "first-order" branches (those connected directly to the apical trunk) to that for the higher-order branches was from 1.41 to 1.73 depending on technique used (see Fig. 4 in \cite{p03}). However, we do not expect that using a distribution of values for $\alpha_i$ would change our conclusions below.

Fig.~\ref{mechSim}(b) illustrates the performance in classifying the 7000 patterns for the three neuron models. The fraction of correct classifications for the purely electrical neuron ($\gamma_j' = \gamma_j'' = 0$) as a function of the global output threshold $x_0$ was at best 64\%. It can be seen that such poor performance stems mainly from the relatively large variance in the overall input intensity $N_e$: although each subunit "tries" to detect the coincidence, background signals from other subunits hinder the detection at the neuron level. The addition of either type of mechanical memory improved the performance dramatically, with more than 90\% correct classification for $\gamma_j' \ge 6$ mV and $\gamma_j'' \ge 11$ mV. 

For the neuron with the global mechanical memory and a large $\gamma_j''$ (e.g., $\gamma_j'' \ge 11$~mV), the stretch activation at the axon initial segment effectively acts in a digital manner, providing a feedback signal that a structural modification has occurred somewhere in the dendritic tree, much like a BPAP is thought to tell the neuron the axon has fired. Several morphological observations favor such memory model for pyramidal neurons. Unlike many other neuron types, the pyramidal neurons have rather straight dendrites that tend to branch at small angles, which should facilitate transmission of the cytoskeletal and cytoplasmic stresses along the dendrite length. The dendritic microtubules are linear, quite rigid and invade the spines, where they likely link to actin cytoskeleton \cite{koro10,hotu10}, which should also facilitate the transmission.

A 1-ms cytoplasmic pressure pulse propagates along a 1-$\mu$m-diameter unmyelinated axon with the velocity 1.1~m/s and decay length 0.18~mm \cite{rvac10}. In the "high viscosity" regime applicable to such pulses, these quantities scale as $\sqrt{\omega d}$ and $\sqrt{\frac{d}{\omega}}$, respectively, where $\omega$ is the central frequency of the wave packet and $d$ is the axon diameter \cite{rvac10} (see also \cite{ciar12}). Assuming that unmyelinated axons and dendrites have similar mechanical properties, the propagation parameters for cytoplasmic pressure pulses should be similar for dendrites. In particular, a 10-ms pressure pulse in a 1-$\mu$m-diameter dendrite should travel with 0.35~m/s velocity and 0.57~mm decay length; for a 100-ms pulse these should change to 0.11~m/s and 1.8~mm, respectively. These values are consistent with the idea that mechanical forces can be transferred through the lengths of the pyramidal neuron dendrites and that the forces can be produced and transmitted sufficiently rapidly so as to be associated with the spike initiating event.

Given these observations, it is suggested that the mechanical mechanism for the combinatorial memory may have evolved relatively recently, culminating in the creation of the pyramidal neuron in higher animals, which allowed the neuron to more specifically respond to the combinatorial aspects of inputs. Note that the mechanical mechanism suggested here confers a functional role to the spine head volume and high spine actin content, roles of which are still enigmatic \cite{spru08,hotu10,kasa10b}.

\subsection{Neuronal circuitry layout}
\label{sec2.3}
\begin{figure}[t]
\begin{center}
\includegraphics[width=\textwidth]{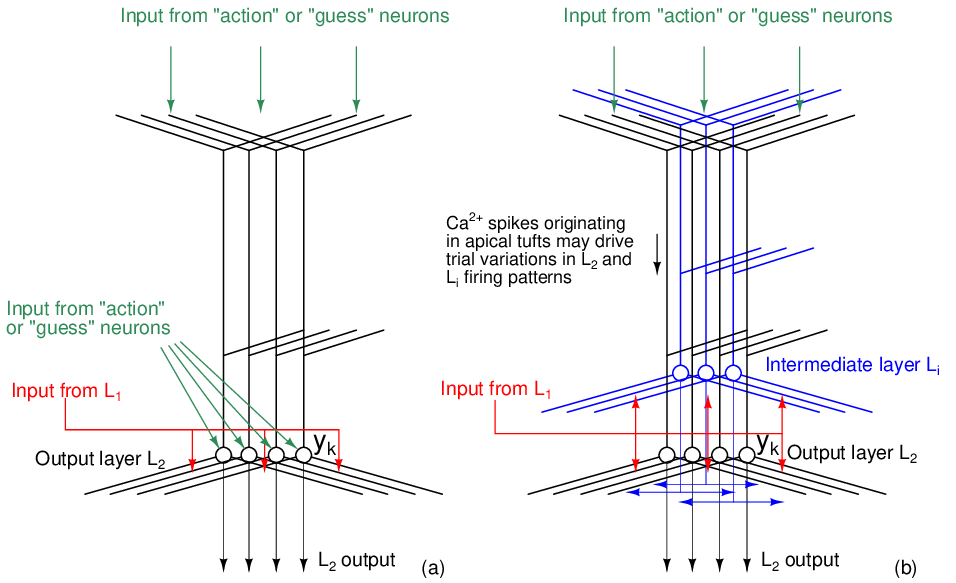}
\end{center}
\caption{\label{figArch} Suggested neuronal architecture for the learning process and memory readout. (a) Learning with one input layer, $L_1$ (red), and one output layer, $L_2$ (black). Input from  $L_1$ diffusely projects into $L_2$ dendrites that are arranged in a plane. The "guess" and "action" inputs (green) connect at the soma or the apical tufts. (b) Learning with an added intermediate layer $L_i$ (blue). Layer $L_1$ diffusely projects into both $L_2$ and $L_i$, while $L_i$ diffusely projects into $L_2$. The "guess" and "action" neurons connect at the apical tufts or the soma (not shown). First, $L_i$ neurons learn to fire for the important to the organism combinations $X$ in $L_1$. The reduced dimensionality signals are then used in further $L_2$ learning. In the neocortex, $L_2$ could correspond to the layer V pyramidal neurons and $L_i$ to the layer II/III pyramidal neurons.}
\end{figure}

Fig.~\ref{figArch}(a) 
shows a suggested neuronal circuitry layout for the learning process and memory readout. Lets assume that in an untrained system presented with an $X$ in $L_1$ the postsynaptic integration does not suffice to excite an $L_2$ neuron $y_k$. In a learning trial, $y_k$ is activated by additional depolarization created by increased excitation or reduced inhibition from one or more "guess" ($G$) or "action" ($A$) neurons that connect to $y_k$ in dominant positions, such as near the axon initial segment. Alternatively, the "guess" or "action" neurons could connect to $y_k$ at the apical tuft, where they could generate the Ca$^{2+}$ dendritic spikes propagating towards the soma and driving initiation of the action potentials \cite{spru08}. The general learning scheme with a global reinforcement signal $R$ broadcast by a critic to all neurons and the neurons receiving "empiric" synapses driven by random spike trains from an external experimenter was first suggested in \cite{seun06}.

Output neurons could be structurally connected to inputs in a similar, although not necessarily identical, manner, such as when closely spaced $L_2$ neurons sprawl basal dendrites in a plane, into which $L_1$ axons diffusely and randomly project (Fig.~\ref{figArch}(a)). This connectivity would be conducive to increasing the learning power of the system, as each $L_2$ neuron would roughly be equal in its ability to learn how to react to an arbitrary combination $X$. In an untrained system, given an $X$ in $L_1$, all $L_2$ neurons should then be similarly close to the activation threshold. Following learning, the $L_2$ neurons trained to react positively to $X$ should be closer to the activation threshold than others. The actual memory readout could proceed using the "action" neurons that deliver similar rising levels of depolarization to all $L_2$ neurons, e.g., via somatic or apical tuft connections, until a certain criterion such as a predefined level of the aggregate $L_2$ output activity is met.

\section{Pattern classification and generalization}
\label{sec3}
We first consider several standard benchmark problems for binary input neural networks and then consider the problem of sparse input classification and generalization.
\subsection{Standard benchmark problems}
\label{sec3.1}
\begin{figure}[t]
\begin{center}
\includegraphics[width=0.5\textwidth]{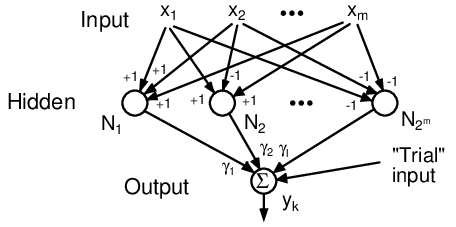}
\end{center}
\caption{\label{figBinClass} Equivalent neural network for a single-neuron solution to the problem of classification of binary patterns $X=\{x_i\}$ randomly assigned to two classes. The hidden computation layer contains units $N_j$, $j=1,\ldots, 2^m$ that correspond to synaptic clusters on the $y_k$ dendrites.  Each unit $N_j$ represents one of the $2^m$ unique patterns $X$ using its $m$ input weights $w_{ji}\in\{-1,1\}$. The unit input-output function is given by $\gamma_j H(n_j-n_j^*-\overline{n}_j)$ with the notation defined in Eq.~\ref{eq1}. Here we assume $n_j^*$ can be equal to 0, i.e., $n_j^* \ge 0$. Output of $y_k$ is the simple sum of its inputs.}
\end{figure}

Fig.~\ref{figBinClass} shows 
the equivalent neural network for a trivial solution to the problem of classification of binary patterns randomly assigned to two classes within the presented framework. $m$ input neurons  form $l=2^{m}$ synaptic clusters of size $m$ synapses each (one synapse per input neuron in each cluster) on the $y_k$ dendrites, with each cluster $j$ representing one of the $2^{m}$ unique patterns $X=\{x_i\}$ using the set of weights $w_{ji}$ from $x_i$ to $N_j$ that take on values of 1 or -1 for an active or inactive $x_i$, respectively, in the pattern $X$ being represented. The cluster input-output functions are given by Eq.~\ref{eq1} and are summed to yield the neuron output
\begin{equation}
\label{eq6}
y=\sum_{j=1}^l\gamma_j H(n_j-n_j^*-\overline{n}_j).
\end{equation}
It can be seen that for every input pattern $X$ only one neighborhood $j$ is "excited", i.e., its $H(n_j-n_j^*-\overline{n}_j)$ is equal to 1, leading to $y = \gamma_j$. Initially, all $\gamma_j$ are set to 0. A weight $\gamma_j$ is set to 1 if its neighborhood $N_j$ is excited, this is followed by a BPAP at $N_j$, i.e., $y_k$ fires, and the following global reward signal $R$ is positive:
\begin{equation}
\label{eq7}
\textrm{Set } \gamma_j = 1 \textrm{ if } H(n_j-n_j^*-\overline{n}_j) = 1 \textrm{, $y_k$ fires and the following } R>0.
\end{equation}

Given a random mapping $\{X_i\} \rightarrow \mathbb{X}^{(0)} , \mathbb{X}^{(1)}$ training proceeds as follows. A pattern $X_i \in \mathbb{X}^{(1)}$ is excited, this is followed by a "trial" $y_k$ firing and the delivery of positive $R$.  The weight $\gamma_j$ whose $N_j$ was excited by $X_i$ is set to 1, in accordance with the learning rule in Eq.~\ref{eq7}. The procedure is repeated for every $X_i \in \mathbb{X}^{(1)}$, resulting in $y_k$ being trained to respond with output 0 to all $X_i \in \mathbb{X}^{(0)}$ and with 1 to all $X_i \in \mathbb{X}^{(1)}$ thus solving the classification problem. Note that this solution requires $2^m(m+1)$ synaptic contacts and at most one training pass or "epoch".

A special case of the above classification problem is the $N$-bit party problem in which the neuron is required to output 1 if the number of 1s in its $N$ inputs is odd, and 0 otherwise. Within the presented framework this is solved similarly, resulting in $\gamma_j$ equal to 1 for those $X_i$ that have odd number of active inputs, and to 0 otherwise. As in the classification problem, the solution requires $2^m(m+1)$ synaptic contacts for $m$ inputs. Table~1 
\begin{table}[t]
\begin{center}
\small
\begin{tabular}{lcccc}
	\hline
Parity& RMCS&  Back-propagation\cite{tesa88}&  EPNet\cite{yao97}& FCEA\cite{yang01}\\
\cline{2-5}
problem    & Number of & Number of& Number of & Number of  \\
        & links (epochs)& links (epochs) & links (epochs)& links (epochs) \\
	\hline
7-bit &1024(1)&127(781)&34.7(177417)&64(1052)\\
8-bit &2304(1)&161(1953)&55(249625)&81(3650)\\
9-bit &5120(1)&N/A&N/A&100(6704)\\
10-bit &11264(1)&N/A&N/A&121(9896)\\
\hline
\end{tabular}
\end{center}
\caption{Comparison of speed and efficiency of solving 7- to 10-bit parity problems in the presented RMCS (Reward-Modulated Combinatorial Switch) model to other models. In the RMCS parity problems are solved exactly in one pass. For $m$ inputs the number of links used is $2^m(m+1)$.  Back-propagation model uses the $N$-$2N$-1 configuration, fully connected from inputs to hiddens and from hiddens to output \cite{tesa88}. EPNet and FCEA are evolutionary algorithms that combine architectural evolution and weights optimization~\cite{yao97,yang01}. "N/A" denotes "Not available".}
\end{table}
compares the speed and efficiency of solving 7- to 10-bit parity problems in the presented model to several other neural network models. The presented model is much more efficient in training time, but much less efficient in the number of synapses used compared to the back-propagation~\cite{tesa88} and evolutionary~\cite{yao97,yang01} models listed. The presented model uses only one, albeit much more intricate, neuron, compared to many more in the other models.

\begin{figure}[t]
\begin{center}
\includegraphics[width=0.88\textwidth]{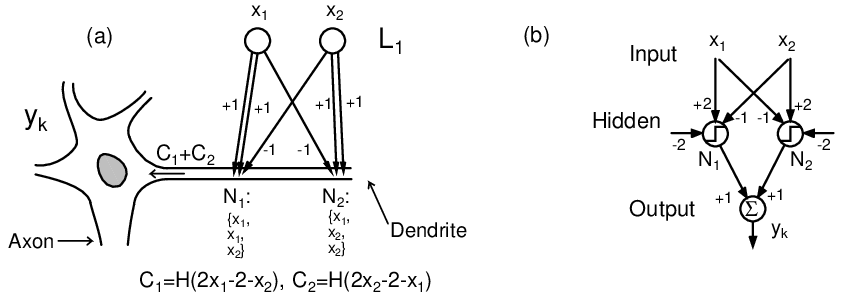}
\end{center}
\caption{\label{FigureXOR} (a) A solution to the XOR problem within the presented framework. The $y_k$ output is the sum of the cluster input-output functions, $C_1=H(2x_1-2-x_2)$ and $C_2=H(2x_2-2-x_1)$. For every binary input pattern $X$ at most one cluster is active. (b) The equivalent neural network diagram for (a).}
\end{figure}

Fig. \ref{FigureXOR} 
shows one of possible solutions to the XOR problem (the 2-bit parity problem) within the presented framework. As in the above solution to the binary pattern classification problem, the cluster excitation thresholds are set to the number of their excitatory synapses $n^*_j$. However, here a cluster may receive more than one synapse from the same input neuron.

\subsection{Sparse coding problems}
\label{sec3.2}
To study how the presented framework performs in the arbitrary $X \rightarrow Y$ mapping task posed in Sec.~\ref{sec1.1} in a sparse coding regime the following simulation setup was created. Here for clarity we modify the notation. $N_i$ input neurons form synaptic clusters on dendrites of $N_o$ output neurons, $N_c$ clusters in total per output neuron, each cluster having exactly $n_c$ excitatory and no inhibitory synapses (the equivalent neural network diagram is similar to the one depicted in Fig.~\ref{figAppleLayout}(a) for a problem considered later). $N_p$ different randomly generated binary input patterns, each having exactly $N_e$ active inputs, are randomly assigned to the  $N_o$ outputs, $N_p/N_o$ patterns per output. The task is to train the system to classify the $N_p$ patterns into the assigned $N_o$ output classes, i.e., make it fire the correct output neuron given presented patterns. The total number of synapses per output neuron, $n_c N_c$, was restricted to $N_s^{(max)}$. 

Training proceeded as follows. The $N_p$ patterns were presented sequentially $t$ times each, with $N_{noise}$ additional random inputs activated at each presentation. Each presentation was followed by a "trial firing" of the correct output neuron. Positive reward $R$ was delivered. Cluster weights $\gamma_j$, initially set at 0, were incremented by 1 for all the synaptic clusters that were excited on the neuron that fired. For this purpose a synaptic cluster was defined as excited if at least $n_{learn}$ ($n_{learn}\le n_c$) its synapses were excited. 

In the testing phase, the same $N_p$ patterns were again presented sequentially, each pattern only once, with additional $N_{noise}$ random inputs activated at each presentation. The output neuron with the largest $\Gamma = \sum \gamma_j$, where the sum is over the clusters that had at least $n_{recall}$ ($n_{recall} \le n_c$) active inputs, was the one that fired. In cases when several neurons had the same largest $\Gamma$ one of them was randomly selected  to fire.

Table 2 shows the results
\begin{table}[t]
\begin{center}
\small
\begin{tabular}{cccccccccccc}
$N_{dupl}$ & $n_c$ & $N_e$ &  &  &  &  &  &  &  &  &  \\ 
\cline{3-12}

& & 1 & 2 & 3 & 4 & 5 & 6 & 7 & 8 & 10 & 15 \\ 
\hline
1 & 1 & 100 & 29 & 19 & 20 & 21 & 20 & 21 & 22 & 19 & 22 \\ 
1 & 2 &  & 55 & 40 & 42 & 42 & 39 & 37 & 38 & 34 & 28 \\ 
1 & 3 &  &  & 67 & 78 & 77 & 75 & 74 & 67 & 59 & 39 \\ 
1 & 4 &  &  &  & 34 & 39 & 49 & 53 & 61 & 63 & 50 \\ 
1 & 5 &  &  &  &  & 33 & 39 & 48 & 59 & 73 & 60 \\ 
1 & 6 &  &  &  &  &  & 31 & 40 & 45 & 58 &  \\ 
0 & 1 & 100 & 29 & 19 & 20 & 21 & 20 & 21 & 22 & 19 & 22 \\ 
0 & 2 &  & 100 & 62 & 57 & 52 & 47 & 44 & 41 & 36 & 29 \\ 
0 & 3 &  &  & 100 & 100 & 99 & 98 & 96 & 89 & 76 & 41 \\ 
0 & 4 &  &  &  & 38 & 79 & 94 & 98 & 99 & 97 & 59 \\ 
0 & 5 &  &  &  &  & 13 & 34 & 73 & 88 & 98 & 83 \\ 
0 & 6 &  &  &  &  &  & 10 & 17 & 34 & 86 & \\ 
\hline
\end{tabular}
\end{center}
\caption{Percentage of patterns classified correctly into $N_o=10$ classes, for $N_p=1000$ different random binary patterns generated on $N_i= 30$ inputs, for various values of the pattern size $N_e$ and the synaptic cluster size $n_c$, for the model of Sec.~\ref{sec3.2}. Clusters were ($N_{dupl}=1$) or were not ($N_{dupl}=0$) allowed to have more than one input from the same neuron. For $N_e = 1,2$ the number of generated patterns was 30 and 30$\cdot$29/2=435, respectively. The limit of $N_s^{(max)}=40000$ synapses per output neuron. For  $n_c = 1,2,3$ and $N_{dupl} = 1$, all $30^{n_c}$ clusters resulting from all possible input permutations were simulated for each output neuron. For  $n_c = 2,3$ and $N_{dupl} = 0$, the $30^{n_c}$ clusters were simulated and those with more than one input from the same neuron eliminated. No random noise added to the patterns. All $n_c$ cluster inputs must be excited for the cluster to learn and to output what it learned ($n_c=n_{learn}=n_{recall}$). Each measurement used a single simulated set of neurons and their interconnections, and the same initial random number seed.}
\end{table}
of the experiment for $N_i=30$, $N_o=10$, $N_p=1000$, $N_s^{(max)}=40000$, $t=1$, $N_{noise}=0$ for different values of $n_c$ and $N_e$. Here we used $n_{learn}=n_{recall}=n_c$, i.e., all $n_c$ cluster inputs were required to be excited for a cluster to learn and to output what it learned. $t=1$ and $N_{noise}=0$ means that there was no randomness introduced during learning and testing, i.e., the task was to simply memorize presented patterns. In one cluster generation method ($N_{dupl}=1$), each synapse was randomly generated from the $N_i$ inputs with equal probability  $1/N_i$. In the second method ($N_{dupl}=0$), after clusters were generated as described above the clusters that had more than one contacting synapse from the same input neuron were eliminated.

From the results in Table 2 it is rather clear that the best performance in the memorization task for pattern sizes $N_e\le 10$ was for the setup with the cluster size $n_c$ of 3 or 4. This can be understood by noting that for $N_i=30$ and with the limitation of $N_s^{(max)}=40000$, the maximum cluster size $n_c$ that would allow the exact representation of all possible patterns of size $N_e=n_c$, i.e., maximum $n_c$ such that  $\frac{n_c N_i!}{(N_i-n_c)!n_c!}\le N_s^{(max)}$, is $n_c=3$. Table~3 shows 
\begin{table}[t]
\begin{center}
\begin{tabular}{lccccccc}
\hline
$n_c$ &  1  &2 &3  &4&5&6&7\\

$N_i$   & 40000 & 200 & 44 & 23&17 &15 &14\\
\hline
\end{tabular} 
\end{center}
\caption{The maximum number of inputs $N_i$ for which all binary patterns of size $N_e=n_c$ may be exactly represented by individual clusters of size $n_c$ using at most 40000 synapses. That is, maximum $N_i$ such that $\frac{N_i!n_c}{(N_i-n_c)!n_c!}\le 40000$.}
\end{table}
the maximum number of inputs $N_i$ for which all patterns of size $N_e=n_c$ can be exactly represented by individual neuron clusters using at most $N_s^{(max)}=40000$ synapses.

Another interesting conclusion from the results in Table~2 is that the elimination of the clusters that had more than one input from the same neuron (i.e., using $N_{dupl}=0$) improves the memorization performance for $n_c = 3$ and 4 dramatically as the clusters representing lower-dimensional pattern features are eliminated.

Table~4 
\begin{table}[t]
\begin{center}
\small
\begin{tabular}{cccccccccccc}
$n_c$ & $t$ & $N_{noise}$ & $n_{learn}$ & $n_{recall}$ & $N_e$ &  &  &  &  &  & \\ 
\cline{6-12}
 &  & & &  & 3 & 4 & 5 & 6 & 7 & 8 & 10 \\ 
 \hline
3 & 1 & 0 & 3 & 3 & 100 & 100 & 99 & 98 & 96 & 89 & 76 \\ 
3 & 1 & 0 & 2 & 2 & 55 & 47 & 41 & 33 & 31 & 31 & 27 \\ 
3 & 1 & 0 & 3 & 2 & 26 & 23 & 24 & 21 & 21 & 21 & 17 \\ 
3 & 3 & 1 & 3 & 3 & 63 & 79 & 80 & 76 & 70 & 64 & 50 \\ 
3 & 3 & 1 & 2 & 2 & 27 & 27 & 26 & 26 & 24 & 26 & 21 \\ 
3 & 3 & 1 & 3 & 2 & 27 & 22 & 22 & 20 & 20 & 21 & 16 \\ 
3 & 3 & 2 & 3 & 3 & 33 & 42 & 47 & 46 & 43 & 42 & 36 \\ 
3 & 3 & 2 & 2 & 2 & 23 & 19 & 22 & 22 & 22 & 22 & 20 \\ 
3 & 3 & 2 & 3 & 2 & 23 & 19 & 21 & 21 & 20 & 20 & 16 \\ 
4 & 1 & 0 & 4 & 4 & 10 & 38 & 79 & 94 & 98 & 99 & 97 \\ 
4 & 1 & 0 & 3 & 3 & 99 & 96 & 87 & 76 & 63 & 51 & 39 \\ 
4 & 1 & 0 & 4 & 3 & 10 & 28 & 30 & 34 & 33 & 27 & 26 \\ 
4 & 3 & 1 & 4 & 4 & 13 & 40 & 68 & 80 & 84 & 87 & 73 \\ 
4 & 3 & 1 & 3 & 3 & 44 & 53 & 44 & 41 & 36 & 31 & 28 \\ 
4 & 3 & 1 & 4 & 3 & 37 & 42 & 36 & 36 & 34 & 28 & 24 \\ 
4 & 3 & 2 & 4 & 4 & 17 & 34 & 46 & 56 & 58 & 57 & 48 \\ 
4 & 3 & 2 & 3 & 3 & 26 & 29 & 29 & 28 & 26 & 23 & 23 \\ 
4 & 3 & 2 & 4 & 3 & 26 & 32 & 32 & 29 & 29 & 24 & 22 \\ 
\hline
\end{tabular}
\end{center}
\caption{Percentage of patterns classified correctly into $N_o=10$ classes, for $N_p=1000$ different random binary patterns generated on $N_i= 30$ inputs, for various values of the pattern size $N_e$ and the synaptic cluster size $n_c$, for the model of Sec.~\ref{sec3.2}. During learning, each of the 1000 patterns was presented $t$ times with $N_{noise}$ additional random inputs activated. During learning, a cluster weight $\gamma_j$ was increased by 1 if the cluster had at least $n_{learn}$ active synapses and its neuron fired. During testing, each of the 1000 patterns was presented once with $N_{noise}$ additional random inputs activated. A cluster weight $\gamma_j$ contributed to the neuron output if the cluster had at least $n_{recall}$ active inputs. $N_s^{(max)}=40000$, $N_{dupl}=0$. For  $n_c = 3$, all $30^{n_c}$ clusters resulting from all possible input permutations were simulated after which those with more than one input from the same neuron eliminated. Each measurement used a single simulated set of neurons and their interconnections, and the same initial random number seed.}
\end{table}
shows the results of experiments in which random noise was added to the patterns during learning and testing. The same base pattern was presented up to three times during learning. In addition, a cluster was allowed to increase its weight and/or contribute to the neuron output when fewer than all its $n_c$ synapses were active ($n_{learn}<n_c$ and $n_{recall}<n_c$, respectively). The conclusion drawn from these experiments is that using $n_{learn}$ and $n_{recall}$ equal to $n_c$ is optimal in most cases. The only situation when using $n_{learn}$ and $n_{recall}$ both lower than $n_c$ helped the classification was when the typical presented pattern was not represented in a cluster but was represented in a subcluster. However, in almost all cases using $n_{learn}=n_{recall}=n_c-1$ was more optimal than using $n_{learn}=n_c$ with $n_{recall}=n_c-1$. That is, the lowering of the cluster excitation threshold after the cluster was trained (as was the case for the function $C(n)$ in Fig.~\ref{figSnScaledDiff}) was typically less optimal than keeping the cluster excitation threshold at a constant, lower or higher, level. 

Another conclusion from Table~4 is that the system exhibited rather high "generalization" performance. For example, the 1000 patterns were classified correctly in 80\% of cases for the cluster size 4 and the pattern size 6 when the patterns were presented 3 times during learning and one additional random input was activated during both learning and testing.

As can be seen from the above simulations, having synaptic clusters that are more specific to the combinatorics of patterns $X$ is desirable for better pattern memorization but this may require a large number of clusters. Lets assume that $n_{learn}=n_{recall}=n_c$. The following question is posed: for $N_i$ input neurons connected to an output neuron $y_k$, $N_e$ the (constant) number of excited neurons in a random binary input pattern $X$, $n_c$ the synaptic cluster size and random $N_i$ to $y_k$ connectivity, how many clusters $N_c$ are needed to fully represent an arbitrary $X$ (without necessarily exactly representing $X$ combinatorics)? The probability that a synapse receives active input from $X$ is $N_e / N_i$. Therefore, approximately, the probability that a cluster is excited, i.e., all its inputs are excited, is $(\frac{N_e}{N_i})^{n_c}$. One needs at least $N_e / n_c$ clusters to represent $X$, leading to $N_c\approx\frac{N_e}{n_c}(\frac{N_i}{N_e})^{n_c}$. Note that, as expected, for $n_c=1$, i.e., nonclustered synapses, this leads to $N_s=n_c N_c = N_i$ synapses needed to represent an arbitrary pattern $X$. For $N_e=n_c$, i.e., exact encoding of $X$ combinatorics in clusters, one needs roughly $n_c N_c={N_e}(\frac{N_i}{N_e})^{N_e}$ synapses. 

Using the above formula for $N_c$ and assuming, for the sake of argument, $N_i=100$ and $N_e=10$, yields $N_s=n_c N_c=10^{n_c+1}$. Comparing this to $50000$, the number of contacting synapses for a pyramidal neuron \cite{spru08}, suggests the cluster size of not more than 3 to 4. Note that for $N_s=50000$, $N_i=100$, $N_e=10$ and $n_c=3$, roughly $N_s(\frac{N_e}{N_i})^{n_c} = 50$ synapses are in the excited clusters for the typical firing pattern.

On the other hand, as discussed in Sec.~\ref{sec6}, most of the learned activity of higher organisms could be considered a form of combinatorial switching if the combinatorial switching idea is taken to its logical extreme. Then, the language, as an artificial human construct designed for ease of communication, should reflect the switching dynamics that the involved neurons are "comfortable" operating on. There are about 40-50 sounds in a typical language and 4-5 sounds in a typical word, which may suggest, roughly, 4-5 as the typical pattern size ($N_e$) and 40-50 as the number of input neurons ($N_i$). Here for simplicity the issue of sound ordering within words is ignored. Using the formulas above with $N_e=5$ and $N_i=50$, the approximate number of contacting synapses per neuron needed to represent an arbitrary word is $5\cdot 10^{n_c}$. Again, comparing this to the experimentally observed 50000 synapses per neuron suggests the cluster size of not more than 4. One could also hypothesize that $N_e$ ($N_i$) can be related to the number of syllables comprising a typical word (the total number of syllables), the number of words in a typical sentence (the total number of words), and the number of the elements in the writing of a typical letter (the total number of the elements).

\section{Training of intermediate layers}
\label{sec4}
To further reduce the dimensionality of inputs through encoding of frequently occurring and significant to the organism input patterns an intermediate layer $L_i$ could be trained using the neuronal architecture suggested in Fig.~\ref{figArch}(b). The axons from $L_1$ randomly project into the basal dendrites of both the intermediate layer $L_i$ and the output layer $L_2$ while the $L_i$ axons randomly project into the $L_2$ basal dendrites. The apical tufts of both $L_i$ and $L_2$ receive the guessing and action driving signals emanating from distal brain areas. Note that the sprawling planar arrangement of $L_i$ and $L_2$ basal dendrites and random synapse connectivity should increase the learning power of the system, as discussed in Sec.~\ref{sec2.3}.

First, $L_2$ neurons are trained to respond to certain $L_1$ combinations. As a side effect, some $L_i$ neurons learn to become more responsive to the frequently occurring $L_1$ patterns that are followed by positive $R$. This learning could occur without the BPAP signals in  $L_i$ neurons if their plasticity can be induced only by the local neighborhood excitation and the subsequent receipt of positive $R$. Then, the trained $L_i$ neurons would become more excited when the learned $L_1$ patterns are presented and could themselves be fired by the guessing or action mechanisms, thus making the reduced dimensionality inputs available to $L_2$ neurons for further learning.

\section{Solution to the multi-neuron combinatorial switching problem}
\label{sec5}
As suggested in Sec.~\ref{sec1}, the organism-level learning problem of finding an optimal $Y^*(X)$ (Fig.~\ref{fig1}(a)) is solved using a trial-and-error search as variations are introduced in the firing combination $Y$. As discussed in Sec.~\ref{sec2.3}, the signals driving the trial variations could come in the form of the Ca$^{2+}$ dendritic spikes originating in the apical tufts. The apical tufts of pyramidal neurons in cortical layers II/III and V are known to receive input from distal cortical areas and nonspecific thalamic projections, with the inputs generally having different origins than those that form synapses with more proximal apical or basal dendrites \cite{spru08}.

It is evident that a proper allocation of behaviors to various $L_2$ neurons or groups of neurons can increase learning efficiency. For example, assume that $L_2$ has $n$ trainable binary-state neurons. Random search for an optimal combination $Y^*(X)$ for a certain $X$, assuming for simplicity that only a single $Y^*(X)$ exists, would consume $\sim 2^n$ trials. This compares to only $n$ trials if one neuron can be trained at a time in any order, or roughly $n(n+1)/2$ trials if one neuron can be trained at a time, but in a particular order that also has to be found by trial and error. The latter training strategies would be possible if $L_2$ neurons drove complementary motor behaviors, such as movements of legs and arms, or rough movements of a leg and finer movements of the leg. The optimal for learning layout of $L_2$ and $L_i$ neurons should certainly be subject to major evolutionary pressures. We have so far considered independent learning for each $y_k$ neuron. However, the excitation of patterns in $L_2$ could be coordinated, e.g., if an $L_i$ neuron drove excitation of several $L_2$ neurons. 

\begin{figure}[t]
\begin{center}
\includegraphics[width=0.95\textwidth]{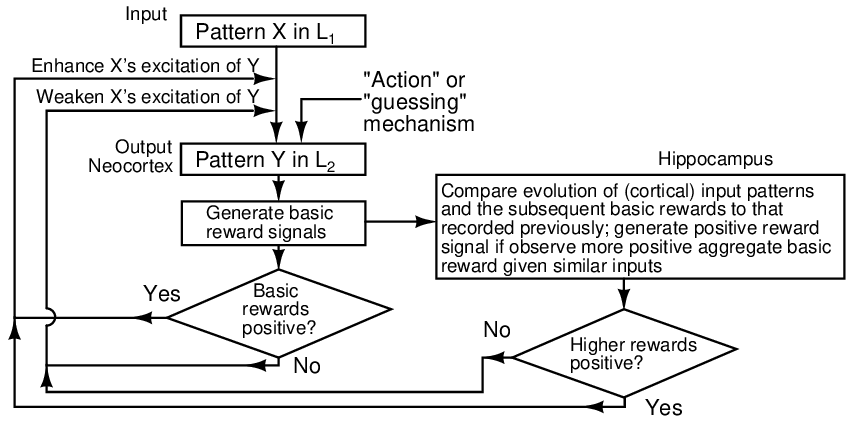}
\end{center}
\caption{\label{figHippocampus} The suggested process of learning in mammals. The diagram is an elaboration of the process illustrated in Fig.~\ref{fig1}(b). The hippocampus plays a role of an "observing body" or "critic" \cite{sutt98} that generates global reward signals. The "basic", or "primary", rewards shown may also be partly generated in the hippocampus. The framework suggests that the higher reward signals may have structurally evolved as an extension of more elementary signals such as pain or hunger.}
\end{figure}

In mammals, it is evident that the "combinatorially trainable" layer $L_2$ and $L_i$ neurons are likely primarily located in the neocortex where they can store complex behaviors. It is suggested that the hippocampus, situated at the edges of the neocortex and indirectly projected into by it, is a major site for generation of basic cognitive and higher global reward signals, or what we suggest may be experienced as "feelings" or "emotions" in humans and higher animals, based on the hippocampus's observation of the neocortical and other brain activity, including the more primary positive and negative rewards generated in other brain regions (Fig.~\ref{figHippocampus}). 
The mechanism of generation of the global reward signal, which probably is an RPE-type signal \cite{bush51a,bush51b,schu97,schu06}, is not the main subject of this paper. However, we note that the mechanism may employ the combinatorial switching principles discussed here to classify synaptic input patterns, although the ability to evaluate the temporal relationships between input signals and gauge the magnitude of the primary rewards would also be needed.

\section{Discussion}
\label{sec6}
Many actions of humans and higher animals seem to fit into the following paradigm: given a combination of sensory inputs, generate an appropriate for the combination action that can be altered through learning. It would be an elegant solution of nature if individual neurons, with some help of auxiliary neuronal circuitry, in fact exhibited this basic behavior---at the single-neuron level expressed as the combinatorial switching of the neuron's output. Indeed, pyramidal neuron connectivity suggests just that: barring necessity for system redundancy, why would a neuron's axon make multiple seemingly randomly distributed connections with another neuron's dendrites, unless there was a combinatorial aspect that is used?

On the other hand, it is widely accepted that higher organisms try to learn to respond to the environment's inputs to achieve positive and avoid negative feelings and emotions \cite{thor11,hilg75,denn81,camp60,czik95}; and that following these subjective learning goals is ultimately connected to the achievement of the organisms' survival and evolutionary objectives.

The idea that emotions play a critical role in learning can be demonstrated with the following example. Consider a toddler learning how to kick a ball to hit a real or imaginary target (creating an implicit, or procedural, memory) by repeatedly kicking the ball and observing its trajectory. What is the mechanism that causes the motor activity associated with more successful trials to be memorized better than that associated with less successful trials, thus allowing the technique improvement? One could suggest that the child consciously and voluntarily, using some mental picture of the process, chooses to remember the movements associated with more successful trials. This would likely require a corresponding cognitive mechanism implemented at the neural level. However, this paper suggests that the positive emotions that accompany the child's realization that an attempt was successful already provide a convenient mechanism for relaying the signal of long-term memorization of the preceding spiking response to the neurons responsible for the more advantageous behavior. Indeed, the reason that emotional responses in humans and higher animals are delivered to a large number (or all, via hormones) of trainable neurons \cite{schu97} may be that the exact site of the neurons being trained, given the complexities of the sensory-motor signal flows, is not easily locatable from the perspective of the emotion generating systems, which themselves may be scattered throughout the nervous system.

An interesting question is: why would the paradigm of combinatorial switching, in which the ability to classify input patterns into output patterns can be considered a multi-neuron implementation, be successful in our world? The answer appears to be that, from a fundamental perspective, the world around us is indeed usefully classifiable, which is in large part driven by the repeating motives in the terrestrial environment and the life organization into similarly behaving species as well as the similarities across the species. (On an even deeper level, these regularities may be viewed as stemming from the invariance of the physical laws in space and time.) A wolf that has learned how to catch a rabbit is more likely to catch another rabbit, as well as another alike animal, in a similar terrestrial environment. The key to efficient learning with a  low-dimensional feedback signal (the reward, or the "emotional response") may be the ability to distil reusable concepts in relatively few learning trials. 

\begin{figure}[thp]
\begin{center}
\includegraphics[width=0.88\textwidth]{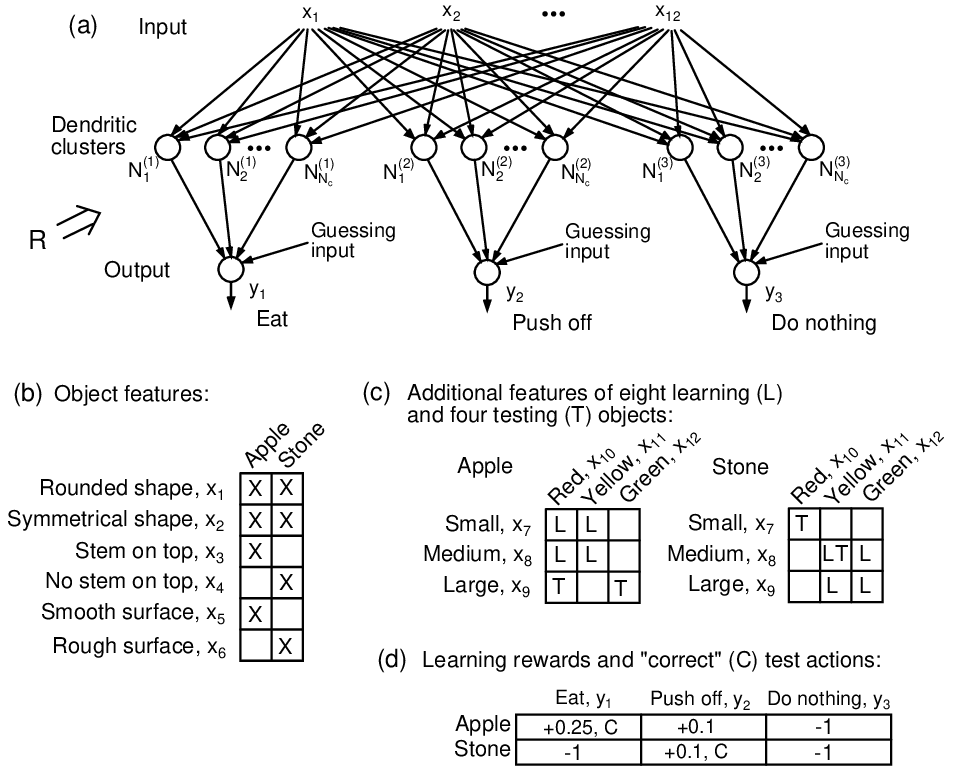}
\end{center}
\caption{\label{figAppleLayout} (a) Equivalent neural network diagram for the problem simulated in Sec.~\ref{sec6}. The 12 binary inputs form $N_c$ clusters of size $n_c$ synapses on the dendrites of each of the 3 output neurons. Each cluster outputs 1 if its weight is at least 1 and its inputs are coincident. Weights are modified during learning. An output neuron $y_k$ fires if the sum of its inputs is at least $M$ or if it receives the "guessing" input. (b) Definition of inputs $x_1, \ldots , x_6$ for the 2 object types: "X" denotes an input of "1", the input is "0" otherwise. (c) Definition of inputs $x_7, \ldots , x_{12}$ for the 8 learning and 4 testing objects. The inputs are "0" or "1" depending on the "L"/"T" position in the matrices. (d) Rewards elicited by the firing of each of the 3 output neurons for the 2 object types. The reward of "-1" resets a cluster weight to 0 if the cluster was excited and its neuron fired. If more than 1 output fires simultaneously the reward of "-1" is generated. "C" denotes the correct output neuron for the purpose of testing.}
\end{figure}

As an illustration of these ideas consider the following simple learning model. An untrained and hungry test subject has 12 sensory neurons connecting to 3 motor neurons (Fig.~\ref{figAppleLayout}(a)). 
All the neurons operate in an "on" or "off" regime. The subject is seated at a table on which apples (rounded symmetrical shape, stem on top, smooth surface) or stones (rounded symmetrical shape, no stem on top, rough surface) are placed one at a time (Fig.~\ref{figAppleLayout}(b)). The apples and stones can be of 1 of 3 sizes (small, medium or large) and 1 of 3 colors (red, yellow or green). Each of the 3 motor neurons drives an action: eating the object on the table, pushing it off the table, or doing nothing, in which case the object is removed from the table following a delay. Each of the sensory neurons fires if its assigned object feature is present: rounded shape, symmetrical shape, stem on top, no stem on top, smooth surface, rough surface, red, yellow or green color, small, medium or large size (the total of 12 features, one feature per sensory neuron).

The sensory neurons connect to the motor neuron dendrites at random locations, forming $N_c$ clusters on each motor neuron, each cluster having $n_c$ excitatory synapses. A cluster is defined as being excited if all $n_c$ its synapses are excited. Each cluster is initially assigned a weight of 0. A neuron fires in a "learned" excitation if at least $M$ its clusters with weights of at least 1 are excited. A weight of 0.25 is added to a cluster for eating an apple, and 0.1 for pushing an object off the table, if 1) all the cluster's synapses are excited, 2) this is immediately followed by a trial firing of the cluster's neuron and 3) this is immediately followed by a positive reward. A cluster's weight is reset to 0 if 1) all the cluster's synapses are excited, 2) this is immediately followed by a trial or nontrial (learned) firing of the cluster's neuron and 3) this is immediately followed by a negative reward (Fig.~\ref{figAppleLayout}(d)). Positive reward is generated for eating an apple or pushing an object off the table. Negative reward is generated for eating a stone, doing nothing, or doing more than one action simultaneously (i.e., at least two motor neurons fire). After an object is placed on the table, the subject attempts to execute a memorized action. If there is no memorized action (i.e., less than $M$ clusters with the weight of at least 1 are excited on each of the motor neurons) a random motor neuron fires in a trial firing.

A computer program RMCLS (Reward-Modulated Combination Learning System) implemented the above learning algorithm. To complicate the problem for the subject and to test its deductive reasoning, no green or large apples and no small or red stones were presented during learning, while green large apples and small red stones were presented during testing. Specifically, the subject was presented with a random sequence of 8 objects: small red apple, small yellow apple, medium red apple, medium yellow apple, medium yellow stone,  medium green stone, large yellow stone, large green stone (Fig.~\ref{figAppleLayout}(c)). After each presentation it was recorded whether the subject would have had correct responses (i.e., eating apples and pushing off stones), if tested, to the 4 test objects: large green apple, large red apple, small red stone, medium yellow stone (Fig.~\ref{figAppleLayout}(c)). In some cases the system was not able to learn responses to all 4 test objects even after a large number of trials.

\begin{figure}[t]
\begin{center}
\begin{tabular}{cc}
    \includegraphics[width=.5\textwidth]{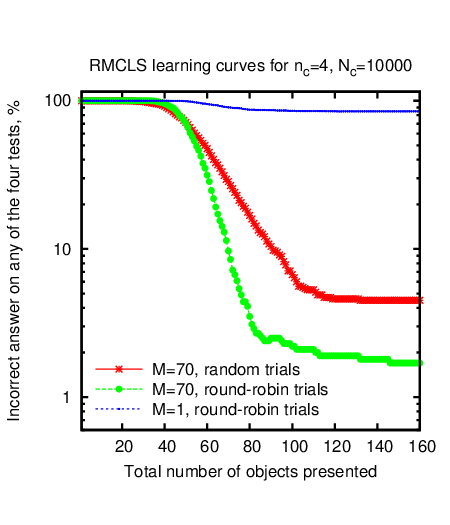} &
   \includegraphics[width=.5\textwidth]{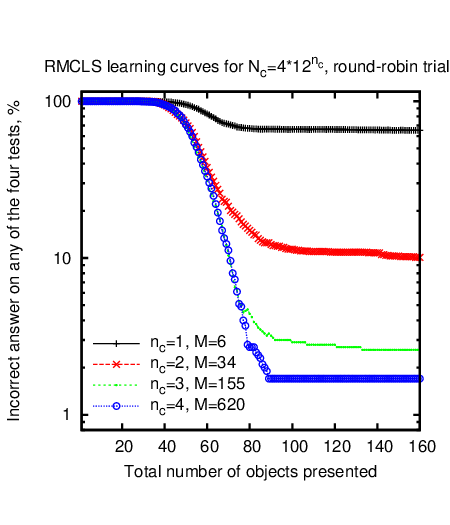}   \\
\end{tabular}
\end{center}
\caption{\label{appleSim} (a) Learning curves for the RMCLS model for $n_c=4$, $N_c=10000$, with the random or round-robin trial firing of the output neurons. The minimum number of excited clusters with the weight at least 1 required to fire a neuron, $M$, was either 70 or 1. Each curve is the average over 1000 statistically independent simulations. (b) Learning curves for the RMCLS model for $n_c=1,2,3,4$ and $M=6,34,155,620$, respectively. $N_c=4\cdot 12^{n_c}$ and round-robin output neuron trials. The values of $M$ were roughly optimal for the learning performance given the values of the other parameters. Each curve is the average over 1000 statistically independent simulations.}
\end{figure}

For $n_c=4$, $N_c=10000$ and $M=70$, the subjects learned to pass all the tests correctly, after a large number of trials, in 95.5\% of cases (Fig.~\ref{appleSim}(a) shows the corresponding learning curve). 
In the other 4.5\% of cases the subjects typically learned the response of pushing both apples and stones off the table. This usually occurred when "pushing off" was, at random, tried many more times than "eating" when an apple was presented; therefore, the subjects had learned to "push off" apples before having tried to "eat" many of them. To make the trial neuron firings more regular the algorithm was modified to select the firing neurons sequentially in a round-robin. Then, the subjects learned to pass the tests correctly in 98.3\% of cases (Fig.~\ref{appleSim}(a) shows the learning curve).

For $n_c=4$, $N_c=10000$ and $M=1$ (with the round-robin motor neuron trials), the subjects learned the 4 correct responses in only 15.3\% of cases. The most common reason for failing a test was due to motor neurons being excited by rarely occurring clusters that represented low-dimensional object features. For example, a cluster with 2 inputs coming from the "rounded shape" sensory neuron and 2 inputs from the "red" sensory neuron would cause all rounded red objects to be classified as edible if the training object sequence happened to have many red apples. Note that out of the $12^4=20736$ clusters representing all possible ordered permutations of 4 out of 12 inputs, 1, 14, 36 and 24 clusters encode the excitation of 1, 2, 3 or 4 particular input neurons, respectively. Therefore, requiring a minimum number of excited clusters to fire a neuron assigned lower importance to one- and two-dimensional object features relative to three- and four-dimensional features.

Next, for each $n_c$ from 1 to 4 (and the round-robin motor neuron trials) the optimal for learning $M$ was searched for, using a large $N_c$, $N_c=4\cdot 12^{n_c}$, so that all possible input combinations were likely to occur in the clusters. For $n_c=1$ the test performance was best when $M$ was equal to 6, with the 4 correct test responses generated in only 34.8\% of cases after a large number of trials; for $n_c=2$, $>$90\% correct responses were obtained for $M$ from 33 to 35 (which represented 5.7-6.1\% of all clusters); for $n_c=3$, $>$95\% correct responses were for $M$ from 115 to 197 (1.6-2.9\% of clusters); and for $n_c=4$, $>$95\% correct responses were for $M$ from 339 to 904 (0.41-1.09\% of clusters). All these measurements were made using 500 statistically independent simulations for each value of $n_c$ and $M$. Clearly, the systems with combinatorial memory ($n_c>1$) performed much better than those without. It is interesting that the range of $M/N_c$ when the test success rate was greater than 95\% was the highest for $n_c=3$. As expected, for low $N_c$ the test performance deteriorated. For example, for $n_c=4$, $N_c=1000$ and $M=7$ the correct responses to the 4 tests were learned in 87.8\% of cases.

Although the RMCLS algorithm is simple, it does suggest that learning in the reward-modulated combinatorial switching framework can be rather efficient, via deduction of reusable abstract concepts. In order to deduce the reusable abstract concepts the system needs to learn in situations that display both these concepts and variability in other features. The system deduces the reusable concepts by accumulating weights for the synapse clusters that represent the concept features. Note that the resulting behavior can be described as "deductive reasoning" and will probably appear intelligent to an external observer. It is evident that in biological neuronal systems the analogues of parameters $n_c$, $M$ and $N_c$ are likely to evolve to suit a particular neuron's operating environment.

The presented framework does not involve value functions and is more alike to policy space search RL algorithms (e.g., evolutionary algorithms~\cite{mori99}) than to value-function-type RL algorithms~\cite{sutt98}. This introduces limitations compared to many well-known value-function RL implementations. Also, it is conceptually possible to apply standard neural network RL algorithms such as the policy gradient method within the presented framework. However, learning through direct interaction with the environment seems more appropriate given that the system tends to have a very large number of weights that are expected to provide a "built-in machinery" for memorization and generalization, and it would probably be difficult to perturb that many weights in a controlled manner while searching in the policy space.

In summary, it is suggested that pyramidal neurons can process information by switching the neuron output based on active input neuron combinations. A trial-and-error learning paradigm is presented in which an (RPE-type) reward signal that itself may adjust over time modulates the combinatorial memory that stores learned behaviors. An experimental verification of the proposed mechanisms, including the putative mechanical or muscle-like contributions that can provide computational advantages to the single-neuron combinatorial switching, is needed.

\section*{Acknowledgments}
The author gratefully thanks Prof. Michael A. Rvachov for prompting exploration of this subject and for many useful discussions. The author also thanks anonymous reviewers for their thorough and helpful reviews.

% Non-BibTeX users please use

\end{document}